\documentclass[pra,twocolumn,nofootinbib,floatfix,10pt]{revtex4-1}
\usepackage[utf8]{inputenc}

\usepackage{amsmath, amsxtra, amssymb, mathtools, isomath, xspace}
\usepackage{bbold}

\usepackage{hhline}
\usepackage{multirow}
\usepackage{sidecap}
\usepackage{graphicx}

\newcommand{\ket}[1]{\ensuremath{\left| #1 \right\rangle}\xspace}

\long\def\symbolfootnote[#1]#2{\begingroup%
\def\thefootnote{\fnsymbol{footnote}}\footnotetext[#1]{#2}\endgroup}

\date{17 May 2016}

\begin{document}

\title{  
Spin and Charge Resolved Quantum Gas Microscopy of Antiferromagnetic Order in Hubbard Chains
}

\author{Martin~Boll$^{1\ast}$}
\author{Timon~A.~Hilker$^{1\ast}$}%
\author{Guillaume~Salomon$^{1\ast}$}%
\author{Ahmed~Omran$^{1}$}%
\author{Jacopo~Nespolo$^{2}$}%
\author{Lode~Pollet$^{2}$}%
\author{Immanuel~Bloch$^{1,2}$}%
\author{Christian~Gross$^{1 \dag}$}%

\affiliation{$^{1}$Max-Planck-Institut f\"{u}r Quantenoptik, 85748 Garching, Germany,}
\affiliation{$^{2}$Fakult\"{a}t f\"{u}r Physik, Ludwig-Maximilians-Universit\"{a}t, 80799 M\"{u}nchen, Germany}

\symbolfootnote[1]{These authors contributed equally to this work.}
\symbolfootnote[2]{Electronic address: {\bf christian.gross@mpq.mpg.de}}





\begin{abstract}
The repulsive Hubbard Hamiltonian is one of the foundational models describing strongly correlated electrons and is believed to capture essential aspects of high temperature superconductivity.
Ultracold fermions in optical lattices allow for the simulation of the Hubbard Hamiltonian with a unique control over kinetic energy, interactions and doping.
A great challenge is to reach the required low entropy and to observe antiferromagnetic spin correlations beyond nearest neighbors, for which quantum gas microscopes are ideal.
Here we report on the direct, single-site resolved detection of antiferromagnetic correlations extending up to three sites in spin-$1/2$ Hubbard chains, which requires an entropy well below $s^*=\ln(2)$.
Finally, the simultaneous detection of spin and density opens the route towards the study of the interplay between magnetic ordering and doping in various dimensions.
\end{abstract}
\maketitle

%
The Hubbard model, describing strongly correlated lattice fermions, supports a rich phase diagram at low temperatures.
Despite the conceptual simplicity of the Hubbard model, parts of its phase diagram, especially away from half filling, and its connection to high temperature superconductivity are still under debate~\cite{anderson1987}.
Here, controlled experiments with ultracold fermions in optical lattices might provide new insight~\cite{hur2009}.
For one particle per site, the so called half filling regime of a balanced two component fermion mixture, and repulsive interactions, the Hubbard model features a crossover from a metallic to a Mott insulating state when lowering the temperature.
For even lower temperatures, antiferromagnetic correlations are expected to develop in the Mott insulating phase due to the superexchange mechanism~\cite{auerbach1994,giamarchi2004,hur2009,esslinger2010}.
The paramagnetic Mott insulating state has been observed in seminal ultracold atom experiments involving trap averaged quantities and, recently, at the single atom level~\cite{joerdens2008,schneider2008,joerdens2010,taie2012,duarte2015,greif2016,cheuk2016}.
Detailed experimental studies of the thermodynamics of the Hubbard model also revealed its equation of state in the density sector down to temperatures at which short range spin ordering might occur~\cite{hofrichter2016,cocchi2016}.
Unfortunately, the experimental preparation of low entropy lattice fermions has proven to be extremely challenging, making the observation of longer ranged antiferromagnetism difficult.
Important progress in revealing magnetic ordering in the Hubbard model has been reported with the observation of nearest neighbor correlations via singlet-triplet spin oscillations~\cite{trotzky2010,greif2013,greif2015} and short range correlations deduced from optical Bragg spectroscopy~\cite{hart2015}.
However, the detection of the onset of magnetic order turned out to be difficult because of the inhomogeneity of the trapped samples, in which different phases coexist.
Microscopic control or detection helps to overcome this limitation and the analogue of antiferromagnetic correlations has been measured in small systems of up to three fermions~\cite{murmann2015a}.
Recently, local, non spin-resolved detection of ultracold fermions in single lattice sites has been demonstrated~\cite{cheuk2015,parsons2015,haller2015,edge2015} and the non-uniform entropy distribution in band and Mott insulating states has been observed in the density sector~\cite{omran2015,greif2016,cheuk2016}.

Here we report on a site- and spin-resolved study of antiferromagnetic correlations in one-dimensional spin-$1/2$ Hubbard chains realized with ultracold lithium-6 in an optical superlattice.
Using our novel spin and onsite atom number sensitive quantum gas microscope, we directly measure spin correlations together with density fluctuations in the system.
The measurements reveal finite-range antiferromagnetic spin correlations extending over up to three sites.
Furthermore, we measured the strength of the spin correlations for increasing interactions, revealing adiabatic cooling at large interaction strengths.
Finally, we observed the decrease of antiferromagnetic correlations away from half filling and the freezing out of density fluctuations at low entropy.

\begin{figure*}[]
\centering
\includegraphics[]{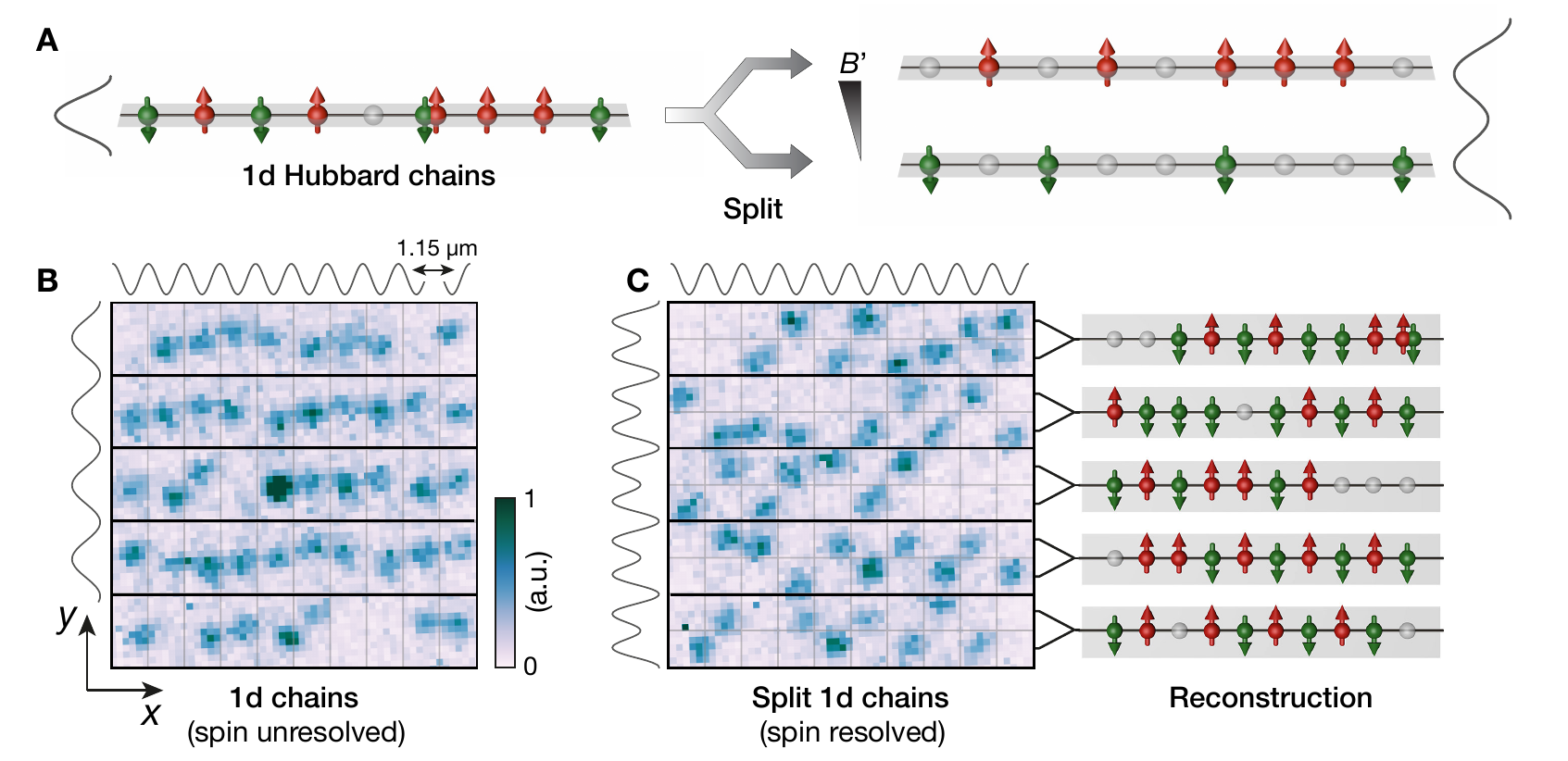}
\caption{\textbf{Schematic of the spin and density resolved detection.}
\textbf{(A)} Schematic of the spin resolved imaging.
Each site of the Hubbard chain was split spin-dependently into a local double well potential.
During the splitting process a magnetic field gradient $B'$ was applied to separate the two spins.
This allows for the simultaneous detection of up spins ($\ket{\uparrow}$, red), down spins ($\ket{\downarrow}$, green), doublons (up and down spins overlapping) and holes (gray spheres) and thus for a full characterization of the Hubbard chains.
\textbf{(B)} Typical fluorescence image of atoms in five mutually independent one-dimensional tubes imaged prior to splitting.
The lattice potentials are indicated by the black lines next to the images with a spacing along the tubes oriented in the $x$-direction of $1.15\,\mu$m and a transverse inter-tube separation of $2.3\,\mu$m.
The increasing fluorescence level is shown by darker colors in relative units as represented by the color bar.
The imaging slightly displaces the atoms from their original positions and also allows for the detection of doubly occupied sites (saturated signal in the center)~\cite{omran2015}.
\textbf{(C)} Typical image with spin resolved detection.
A superlattice in the $y$-direction (indicated on the left of the image) was used to split each chain in a spin dependent manner.
The $\ket{\downarrow}$ spins were pulled down, while the $\ket{\uparrow}$ spins were pulled upwards.
The right image illustrates the reconstructed Hubbard chains.}
\end{figure*}


The fermionic atoms in each of the one-dimensional lattice tubes are well described by the single band Hubbard Hamiltonian
\begin{equation}
\hat{H} = -t \sum_{i,\sigma} (\hat{c}^\dagger_{i,\sigma}\hat{c}_{i+1,\sigma} + \mathrm{h.c.})
          +U \sum_i \hat{n}_{i,\uparrow}\hat{n}_{i,\downarrow}
          + \sum_{i,\sigma} \epsilon_{i} \hat{n}_{i,\sigma}\,.
\end{equation}
Here the fermion creation (annihilation) operator is denoted by $\hat{c}^\dagger_{i,\sigma}$ ($\hat{c}_{i,\sigma}$) at site $i$ for each of the two spin states $\sigma = \uparrow, \downarrow$.
The operator $\hat{n}_{i,\sigma}=\hat{c}^\dagger_{i,\sigma}\hat{c}_{i,\sigma}$ counts the number of atoms with spin $\sigma$ on the respective site.
Three competing energy scales govern the physics of this system: intersite nearest-neighbor hopping with strength $t$, onsite interactions of strength $U$ and local trap induced energy offsets $\epsilon_i$.
While $t$ is controlled via the lattice depth, $U$ can be tuned independently in the experiment using the broad Feshbach resonance of lithium-6 between the lowest hyperfine states $\ket{\downarrow} = \ket{F = 1/2,m_F = -1/2}$ and $\ket{\uparrow} = \ket{1/2, 1/2}$~\cite{zuern2013a}.
In the experiments reported here, we exclusively worked with repulsive interactions $U>0$, for which the Hubbard model supports finite range antiferromagnetism with correlations suddenly appearing at distances beyond nearest neighbors for entropies per particle below the ``critical'' value of $s^* = S/N k_B = \ln(2)$~\cite{sirker2002, gorelik2012, sciolla2013}.
Importantly, true long range order is absent in the 1d Hubbard model even at zero temperature~\cite{giamarchi2004,essler2005}, and the resulting algebraic decay of the correlations is significant even on a distance of a few sites~\cite{gorelik2012, sciolla2013}.
In the limit of very strong repulsive interactions and half-filling, the emerging spin order is intuitively understood from the mapping of the Hubbard model to a Heisenberg antiferromagnet with superexchange coupling $J=4t^2/U$~\cite{auerbach1994}.
For lower interactions, particle-hole fluctuations become important and the ground state is characterized by a spin density wave. 
In one dimension, the model is Bethe ansatz integrable~\cite{giamarchi2004,essler2005} and precise predictions for the finite entropy spin correlations and density fluctuations have been reported in the relevant parameter regime of cold atom based experiments~\cite{gorelik2012,sciolla2013}.\\


The experiments started with the preparation of a low temperature balanced spin mixture of the $\ket{\uparrow}$ and $\ket{\downarrow}$ states in a single two dimensional lattice plane~\cite{omran2015}.
The final temperature and atom number was controlled by magnetic field driven spill-out evaporation at repulsive interactions~\cite{som}.
We set the final interaction strength using a homogeneous magnetic offset field to control the scattering length in the vicinity of the Feshbach resonance centered at $832\,$G~\cite{zuern2013a}.
Afterwards, we ramped up the large spacing component ($a_{sl} = 2.3\,\mu$m) of a superlattice~\cite{sebby-strabley2006,foelling2007} in the $y$-direction to prepare independent one-dimensional (1d) tubes.
Next, we slowly turned on a lattice with spacing $a_l = 1.15\,\mu$m along the tubes in the $x$-direction using a $100\,$ms linear ramp to $11\,E_R$, where  $E_R=h^2 / 8 m a_l^2$ denotes the recoil energy of the lattice for atoms of mass $m$.
The hopping strength is $t = h \times 125(9)\,$Hz at this final lattice depth.
The lattice filling was controlled by varying the evaporation parameters.
To simultaneously detect the spin and density degrees of freedom of the 1d Hubbard chains locally, we froze the dynamics by rapidly increasing the lattice depth along the tubes to $42\,E_R$ within $1\,$ms, followed by a turn-off of the magnetic offset field in $20\,$ms.
Spin resolution was obtained using the superlattice potential and a magnetic field gradient in the $y$-direction in a Stern-Gerlach like setting.
The magnetic field gradient shifted the potential minima experienced by the two spin states of opposite magnetic moment and the subsequent adiabatic ramp-up of the short scale component of the $y$-superlattice with well separation $a_l$ caused a separation of the spins into the two different sites of the local double well (see Fig.~1a).
Applying this technique to a spin polarized gas, we inferred a splitting fidelity of $98\%$ limited by superlattice phase fluctuations of $25\,$mrad~\cite{som}.
Finally, we ramped up a three-dimensional pinning lattice for detection and reconstructed the lattice site occupations from fluorescence images (see Fig.~1(b,c))  after deconvolution with the measured point-spread-function ~\cite{omran2015,som}. The above detection procedure enables us to detect the position of all spins, doublons and holes in the system with single lattice site resolution, thereby obtaining complete information about the system in Fock space.

\begin{figure}[t]
\centering
\includegraphics[width=\columnwidth]{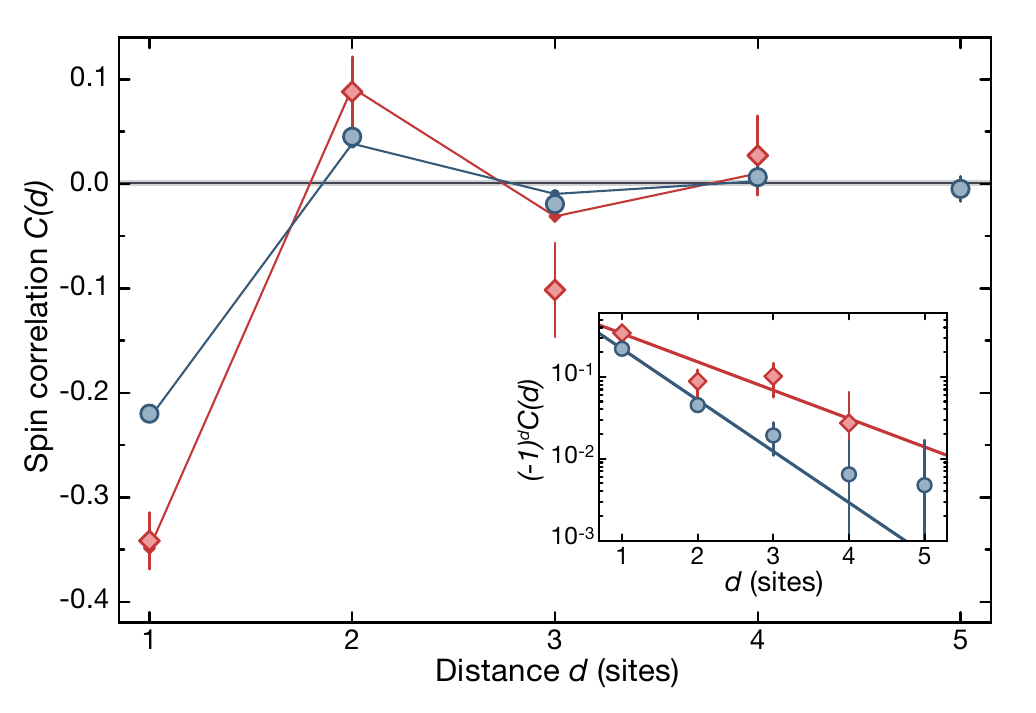}
\caption{
\textbf{Antiferromagnetic spin correlations versus distance.}
The main figure shows the measured spin correlations at $U/t=12.6$ for the loosely (blue circles) and more tightly filtered data (red diamonds), see main text.
The staggered behavior directly visualizes the antiferromagnetic nature of the correlations $C(d)$.
Correlations up to three sites are statistically significant.
The transverse correlations (gray line) vanish within its one s.e.m. uncertainty (light gray shading).
The red and blue lines connecting filled symbols are QMC results for a homogeneous system at half filling corresponding to entropies per particle of $s=0.51(5)$ and  $s=0.61(1)$ respectively.
The inset shows the decay of the staggered spin correlator $C_{s}(d)=(-1)^d C(d)$ in a logarithmic plot together with an exponential fit  $C_s(d)\propto \exp(-d/\xi)$ revealing decay lengths of $\xi = 0.69(6)$ sites and $\xi = 1.3(4)$ sites for the two data sets. For low entropies, an exponential decay is expected to be strictly valid only at large distances. However, within the statistical uncertainty of the experimental data the fit captures the observed behavior well.
All error bars represent one s.e.m.
}
\end{figure}

\begin{figure}[h!!!]
\centering
\includegraphics[width=\columnwidth]{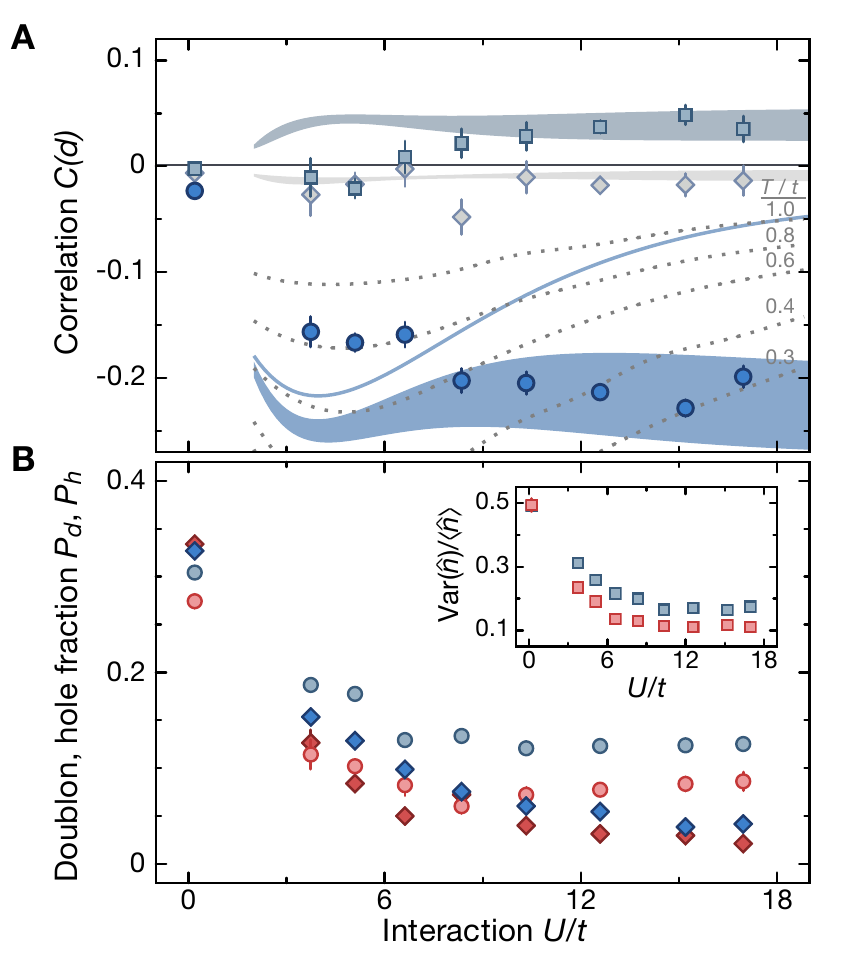}
\caption{
\textbf{Spin and density degrees of freedom at different interaction strength.} 
\textbf{(A)} Spin correlations $C(d)$ for distances $d=1$ (dark blue), $2$ (light blue) and $3$ (gray) versus interaction strength $U/t$.
Starting close to zero at vanishing interactions, finite range spin correlations develop and saturate for interaction strengths $U/t > 8$.
The shaded areas indicate the QMC predictions in a homogeneous system at half-filling for an entropy per particle between $s=0.60$ (lower bound) and $0.65$ (upper bound), the solid line is the prediction for $C(1)$ at $s^*=\ln(2)$. Dotted lines are isothermals for $C(1)$ at the indicated temperature.
For large $U/t$, we observe adiabatic cooling, while both temperature and entropy decrease in the analyzed spatial region at intermediate $U/t$.
The transverse nearest neighbor correlations (dark gray line close to zero) is consistently above the $d=3$ spin correlator, supporting its statistical significance.
Due to limited statistics, only loosely filtered data (see main text) is shown.
\textbf{(B)} Evolution of the density degree of freedom.
The main figure shows the evolution of holes (circles) and doublons (diamonds) with interaction strength $U/t$.
The hole ($P_h$) and doublon ($P_d$) fractions decrease for low interactions and then saturate.
Here, data is shown for the loose (blue) and tight (red) filter case.
The inset shows the evolution of the normalized onsite atom number variance $\mathrm{Var}(\hat{n})/\langle \hat{n} \rangle$.
The density fluctuations are suppressed already at vanishing interactions due to effects of Pauli blocking in the metal.
This suppression becomes stronger for increasing interactions until the fluctuations saturate.
All error bars represent one s.e.m and the apparent fluctuation of the data is due to day-to-day systematics.
}
\end{figure}

First, we analyzed the  spin correlations $C(d) = 4(\langle \hat{S}^z_i \hat{S}^z_{i+d} \rangle - \langle \hat{S}^z_i \rangle \langle \hat{S}^z_{i+d}\rangle)$ between the spin operators $S^z_i=(\hat{n}_{i,\uparrow}-\hat{n}_{i,\downarrow})/2$ versus distance $d$.
To this end, we fixed the $s$-wave scattering length to $671(10)\,a_B$, where $a_B$ is the Bohr radius, corresponding to $U/t = 12.6$ and took a high statistics dataset of $1200$ individual pictures.
We focused on the central region of the inhomogeneous sample, defining two spatial regions of interest for the analysis. The first one, referred to as the  loose filter, involves all sites with an average density $\langle \hat{n}_i \rangle = \langle \hat{n}_{i,\uparrow} + \hat{n}_{i,\downarrow} \rangle$ in the range $\langle \hat{n}_i \rangle = 1\pm0.3$. 
Here, we benefit from higher statistics, but there is a considerable effect of lower density regions in the data.
The second, so called tight filter, selects one specific tube and takes only sites closer to unity density with $\langle \hat{n}_i \rangle = 1\pm0.1$ into account~\cite{som}.
The strong nearest-neighbor correlations of $C(1)=-0.220(5)$ observed for the loose filtering correspond to $37\%$ of the expected zero temperature signal in the Heisenberg limit~\cite{gorelik2012,sciolla2013} (see Fig.~2). 
For the tight filter we even measured a nearest-neighbor correlation of $C(1)=-0.34(9)$ corresponding to $58\%$ of the zero temperature prediction.
This data is based on the average over all measurements taken for $U/t>8$ (cf. Fig.~3).
We observed significant correlations over a distance of up to three sites of the  staggered correlator $C_s(d)=(-1)^d C(d)$.
A comparison between the experimentally measured correlator $C(d)$ and finite temperature quantum Monte Carlo (QMC) calculations for homogeneous Hubbard chains at half filling allows to determine the entropy and temperature of the lattice gas~\cite{som}.
We inferred an effective local entropy per particle of $s=0.61(1)$ in the loosely filtered case, reducing to $s=0.51(5)$ for the tight filter, both, significantly below $s^* = \ln(2) \approx 0.69$. In a uniform system at half filling this lowest entropy corresponds to a temperature of $ k_{\mathrm{B}}T/t = 0.22(4)$ at $U/t = 12.6$~\cite{som}.

In order to explore the properties of the Hubbard chains at different interaction strengths $U/t$, we measured spin correlations and particle-hole fluctuations for varying onsite interactions $U$, while keeping the lattice ramp and final lattice depth constant at $11\,E_R$ (Fig.~3). We compare the measurements to QMC results for a homogeneous system at half filling for different temperatures and entropies.
The dependence of the correlations on the interactions is rather different for isothermal or isentropic state preparation.
In the former case, a maximum of the correlations is expected at intermediate interactions $U/t$, where part of the entropy is carried by density modes~\cite{werner2005}, while at large interactions the correlation decrease due to the smaller energy scale of spin excitations given by the superexchange coupling $J$.
In the isentropic case of constant entropy, spin correlations saturate towards strong interactions, where the energetic gap between spin and density modes is large. At intermediate interaction strengths, the correlation behavior depends on the entropy and a weakly pronounced maximum exists for intermediate entropies around $s^*=\ln(2)$ (cf. Fig.~3a), while below $s=0.6$ a monotone increase of the correlations with interaction strength is expected.
Experimentally, we observed a saturation behavior of the spin correlations for $U/t > 8$. The inferred temperature dropped from $k_B T = 0.6\,t$ to $0.3\,t$ while increasing $U/t$ from $8$ to $16$ as expected for adiabatic cooling.
At intermediate interactions, $U/t \approx5$, we observed reduced spin correlations compared to the isentropic prediction at half filling. We attribute this to a changing entropy distribution in the trap \cite{som, joerdens2010} and a weak increase of the mean density in the analyzed region by $5\%$.
In the regime of saturated spin correlations, the doublon and hole fractions reached their lowest value of $P_d = 5\%$ and $P_h = 12\%$.
The higher hole fraction is mainly due to the loose filtering resulting in a slight effective hole doping in the analyzed region of the system, which is lower ($P_d = 3\%$, $P_h=7\%$) for the tightly filtered data.

\begin{figure}[t]
\centering
\includegraphics[width=\columnwidth]{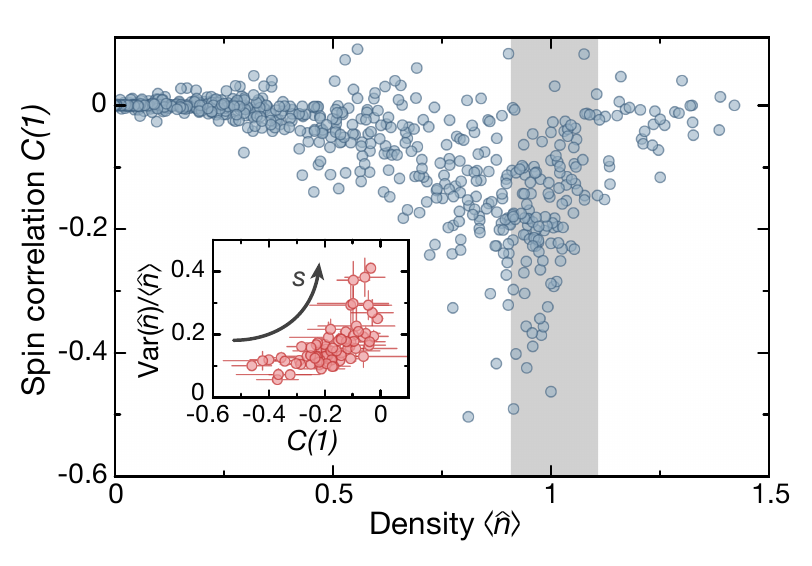}
\caption{
\textbf{Interplay of density and spin fluctuations.}
We show the nearest neighbor spin correlations $C(1)$ for different densities corresponding to different positions in the trap.
The data combines several measurements at an interaction strength of $U/t=10.3$, also including higher temperature data.
Every data point corresponds to two neighboring sites, where between $30$ and $2000$ samples contribute.
The spin correlations $C(1)$ peak just below densities of one, consistent with the half filling regime taking the detection efficiency of about $95\%$ into account.
The inset shows the normalized density fluctuations $\mathrm{Var}(\hat{n})/\langle \hat{n}\rangle$ versus $C(1)$ for a density interval $\langle \hat{n}_i \rangle = 1\pm0.1$ as indicated by the gray area.
Density fluctuations rise steeply for low values of the spin correlator signaling the saturation of the entropy in the spin sector.
This characteristic dependence identifies the strong vertical scatter of the data in the main figure mainly as a result of different local entropies $s$ as indicated by the arrow.
For clarity of the presentation we omit the error bars in the main figure. They are of the same size as the ones in the inset and correspond to one s.e.m.
}
\end{figure}

More insight into the behavior of the spin correlations can be obtained by making use of the full microscopic characterization of the system.
To study the antiferromagnetic spin correlations away from half filling, we show the nearest neighbor correlator $C(1)$ per pair of sites versus their mean density in Fig.~4.
This data combines different datasets taken at $U/t=10.3$ and also contains measurements at different temperatures, obtained by holding the cloud for up to $2.5\,$s in the two-dimensional plane.
We observe a clear dependence of the spin correlator on the local density, with strongest correlations close to $\langle \hat{n} \rangle=1$.
Away from half filling, both to higher and lower densities, a strong decrease of the correlations is observed reflecting the fact that doping reduces spin order~\cite{hur2009}.
Generally, the data scatters much stronger than expected just by statistics, that is, at a given density we observe events with a range of significantly different nearest neighbor spin correlations.
This reflects the distribution of entropy within each cloud, as well as between the measurement settings.
To further analyze the data, we selected a density interval $\langle \hat{n}_i \rangle = 1\pm0.1$ and calculated the normalized variance of the density $\mathrm{Var}(\hat{n})/\langle \hat{n}\rangle$ for all pairs of sites in this window.
These fluctuations reflect the entropy in the density sector, while the nearest neighbor spin correlations are a measure of the spin entropy. We show their mutual dependence in the inset of Fig.~4, identifying two distinct regimes of total entropy.
In the regime below $s^*=\ln(2)$ ($C(1) \lesssim 0.15$), the density fluctuations depend only weakly on the total entropy~\cite{gorelik2012, sciolla2013}, which in turn is stored in the spin fluctuations.
Only when these are saturated at $s^*$, the density fluctuations grow, visible in their steep rise when the spin correlations are just below zero.
The freezing of density fluctuations renders them useless as a thermometer in the low entropy regime, while the highly temperature (and entropy) sensitive spin correlations are ideal for this purpose down to $T=0$~\cite{gorelik2012}. Combining the spin correlation measurements for several distances can further improve such a thermometer \cite{som}.

In conclusion, we have explored the low entropy regime of one-dimensional Hubbard chains in which finite range antiferromagnetism starts to develop.
The inferred local entropy is consistently below the critical value of $\ln(2)$ reaching $s\approx0.5$ in the lowest entropy 1d tube.
Combining our quantum gas microscope with optical superlattices, we demonstrated the simultaneous detection of all relevant degrees of freedom.
We characterized the state of the Hubbard chains at different interaction strength in terms of spin correlations and density fluctuations.
The measurable spin correlation signal extended over three sites, well beyond nearest-neighbors.
Using a local analysis, we were able to present a first study of the behavior of the spin correlations away from half filling.
The demonstrated ability to characterize spin correlations in 1d systems locally provides a useful thermometer in the low entropy regime where density fluctuations are frozen~\cite{gorelik2012,sciolla2013} and cannot serve anymore as a thermometer.
Such a ``spin thermometer''~\cite{olf2015} is a crucial step towards optimized cooling~\cite{bernier2009,ho2009} to even lower entropies required to study, for example, d-wave superfluidity away from half filling~\cite{rey2009}.
Furthermore, the combination of superlattices and local detection will allow for the search of an adiabatic path between low entropy valence bond solids~\cite{trotzky2010} or plaquette resonating valence bond states \cite{trebst2006,nascimbene2012} and the Heisenberg antiferromagnet~\cite{lubasch2011}, also in two dimensions.
Realization of the paradigmatic quantum phase transition from such an artificial valence bond solid to a Heisenberg antiferromagnet~\cite{senthil2004} thereby seem within reach of present experiments.


Recently, we became aware of similar experimental results in two dimensions~\cite{parsons2016, cheuk2016a}.


\clearpage

\section*{Supporting Material}
\section{Cloud preparation}

The experiments started with a degenerate, incoherent spin mixture of the lowest two hyperfine states of lithium-6 in a single two-dimensional plane of a vertical lattice with spacing $a_z=3.1\,\mu$m and depth of $V_z=185\,E_R$.
The resulting harmonic confinement in the $z$-direction was $\omega_z = 2\pi \times 22.5\,$kHz.
Distinct to prior experiments~\cite{omran2015}, this single plane was directly loaded from a strongly elliptical shaped dimple trap propagating in the $y$-direction with beam waists $w_x=10.3\,\mu$m and $w_z=1.7\,\mu$m in the $x$- and $z$-direction, respectively, at a magnetic field of $599\,$G corresponding to a scattering length  $a_s=353\,a_B$.
A further crossed dipole trap beam propagating along the $z$-direction ($w_{xy}=75\,\mu$m) was used to provide additional radial confinement in the two-dimensional plane.
The final evaporation was performed in the single lattice plane with the additional cross trap by ramping up a magnetic field gradient along the $y$-direction in $2.5\,$s to $25\,$G/cm.
After ramping down the magnetic field gradient, we set the scattering length (and thus the final interaction strength $U$ in the lattice) by varying  the homogeneous magnetic field between $529\,$G and $657\,$G .
We adjusted the peak density of the two dimensional gas to approximately $1/(1.15 \times 2.3)\,\mu$m$^{-2}$, the required density for unity lattice occupation, by lowering the power of the cross trap beam.
To prepare one dimensional tubes, we linearly ramped up the long scale component of a superlattice in the $y$-direction ($a_{\rm sl}= 2.3\,\mu$m) to a depth of $34\,E_R$  in $100\,$ms.
The remaining tunnel coupling between the tubes is below one Hertz and negligible on the timescales studied here, yielding decoupled and independent tubes.
Subsequently, we ramped up a short scale lattice ($a_{l}=1.15\,\mu$m) along the tubes in the $x$-direction in $100\,$ms to a lattice depth of $11.0(4)\,E_R$, which results in a tunnel coupling of $t = h \times125(9)\,$Hz.
The lattice depths in the $x,y,z$-direction are given in units of the respective recoil energies: $E_R = h\times6.28\,$kHz, $h\times1.57\,$kHz, $h\times0.87\,$kHz.

\renewcommand{\thefigure}{S\arabic{figure}}
\setcounter{figure}{0}

\begin{figure}[ht]
\centering
\includegraphics[]{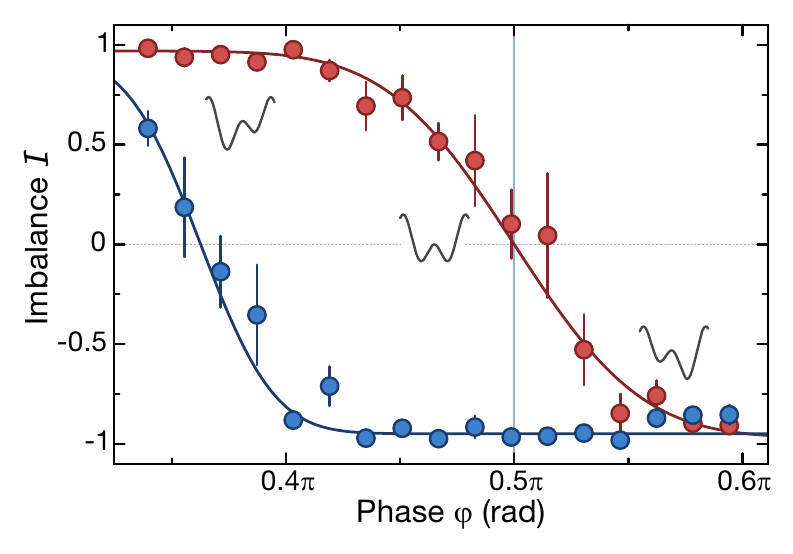}
\caption{
\textbf{Spin selective imaging via deterministic splitting in local double wells.}
The figure shows the local atom number imbalance $I=(n_{\rm left}-n_{\rm right})/(n_{\rm left}+n_{\rm right})$ in the local double wells for the central region of the cloud versus the superlattice phase $\varphi$.
This externally controlled phase $\varphi$ sets the symmetry of the double wells.
For this measurement a spin polarized sample was used, of which the preparation is described in ref.~\cite{omran2015}.
The red data were taken without, the blue data with the magnetic gradient field of $60(5)\,$G/cm, and the solid lines are error function fits to the data.
The splitting was done at the symmetric point ($\varphi=\pi/2$) of the superlattice potential, where the gradient-free measurement showed zero imbalance (gray line).
Taking into account the detection fidelity of $95\%$, the splitting fidelity of $98\%$ is estimated from the value of the imbalance in presence of the magnetic field gradient at $\varphi=\pi/2$.
Error bars denote one s.e.m.
}
\end{figure}

\begin{figure*}[t]
\centering
\includegraphics[]{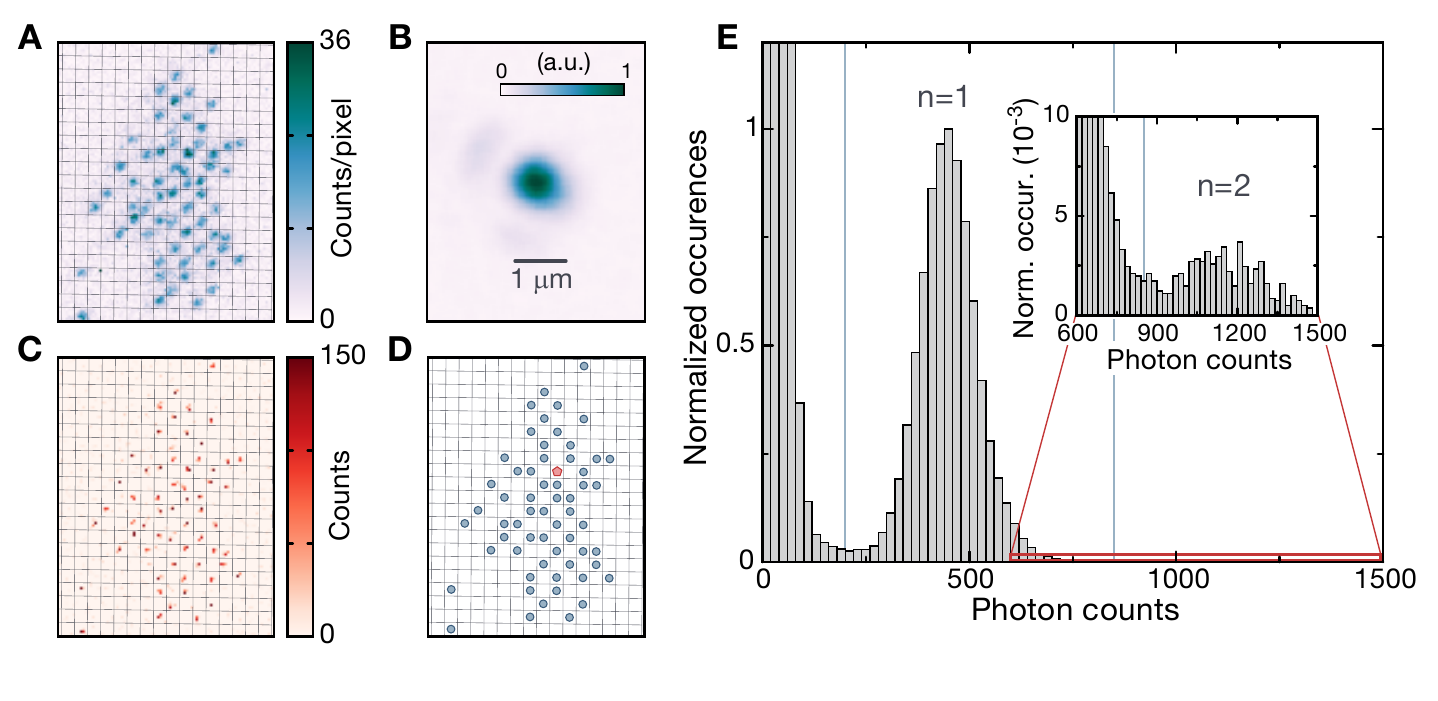}
\caption{
\textbf{Reconstruction of the lattice occupation.}
\textbf{(A)} Typical raw fluorescence image with overlaid lattice grid. The color scale gives the photon counts per pixel of the CCD camera.
\textbf{(B)} Point-spread-function of the imaging system with color scale in relative units. The Gaussian widths are $\sigma_1=314\,$nm and $\sigma_2=363\,$nm along the principal axes.
\textbf{(C)} Deconvolved image from (A) using the point-spread-function shown in (B) using a Richardson-Lucy deconvolution algorithm~\cite{Biggs1997}. The color scale represents the counts per pixel after this deconvolution.
\textbf{(D)} Reconstructed lattice site occupation showing empty sites, singly occupied ones (blue) and one doublon (red).
\textbf{(E)} Histogram of photon counts per site normalized to the $n=1$ peak. The separation between empty and singly occupied sites is excellent, with the counts in the intermediate region dominated by hopping events. The inset is a zoom into the region indicated by the red box revealing the doublon peak, which is clearly visible even for the very low number of detected doublons after magnetic splitting ($<1\%$). We indicate the boundaries for identifying one and two atoms per site by the vertical lines.
}
\end{figure*}

\section{Detection}

\subsection{Spin resolved detection}

The detection sequence was initiated by freezing out the dynamics along the tubes with an abrupt ($1\,$ms) increase of the lattice depths to $42\,E_R$,  $84\,E_R$ and $278\,E_R$  in the $x$- , $y$- and $z$- direction.
Then, we ramped down the magnetic Feshbach field to $1\,$G to enter the Zeeman regime, where both spin components have equal and opposite magnetic moments.
To directly probe both the local population and the spin state, we used the short scale component of the superlattice in the $y$-direction in combination with a magnetic field gradient ~\cite{trotzky2008}. Here, the full superlattice potential is described by
\begin{equation}
V = V_{l} \cos^2 \left( k_l  y + \varphi \right ) + V_{\rm sl} \cos^2 \left (\frac{k_l }{2} y \right)\,,
\end{equation}
with lattice depths $V_{l,sl}$ and wave vector $k_l=\pi/a_l$.
The relative phase of the superlattice $\varphi$ controlling its symmetry was set such that the local double wells were symmetric.
This was verified by measuring the double well imbalance versus superlattice phase $\varphi$ for a spin polarized sample as shown in Fig.~S1.
The magnetic field gradient was first set to $60(5)\,$G/cm before adiabatically ramping up the short scale superlattice component with spacing $a_l = a_{\rm sl}/2$ to $V_{l}= 17\,E_R$ in $10\,$ms while ramping down the long scale component to $V_{\rm sl}= 10\,E_R$. The corresponding energy offset due to the Zeeman shift between the left and the right lattice wells was $\Delta E_{LR} = \pm h\times 6.2(6)\,$kHz for the spin up and down value respectively.
As a result of their opposite magnetic moments, the spin components were adiabatically transferred into different sites of the local double well in a deterministic way and measuring their positions gave access to their spin state.
This sequence ensured a spin separation fidelity of $0.98$ (see Fig.~S1).
For the magnetic field gradient of $60(5)\,$G/cm, we estimated that the relative phase $\varphi=\pi/2$ between the short and long scale components of the superlattice had to be controlled to better than $100\,$mrad, which we extracted from the width of the red curve of Fig.~S1 obtained without applying any magnetic field gradient. The shot-to-shot rms-fluctuation of the phase was measured to be $25 \,$mrad by direct imaging of the superlattice potential after the chamber.

\subsection{Reconstruction of the site occupation}
Compared to our previous publication~\cite{omran2015}, we improved the population reconstruction algorithm.
In this work, we used an accelerated Richardson-Lucy deconvolution~\cite{Biggs1997} with a measured point-spread-function that we obtained by averaging the image of $1000$ isolated atoms.
The resulting deconvolved image after $100$ steps of the Richardson-Lucy algorithm contained the signal for every atom almost on a single pixel.
Integration of the counts in the region of each lattice site allowed us to identify sites with zero, one and two atoms with high fidelity thanks to the parity-free detection~\cite{omran2015}. We obtained a fidelity of $>99\%$ for distinguishing zero and one, and a fidelity of $>95\%$ for distinguishing one and two atoms per site.
Doubly occupied sites after spin splitting arise from atoms in higher band, non perfect splitting and tunneling during the exposure. We detected less than $0.5$ of these events per image for the measurements reported here. The reconstruction of a typical image is summarized in Fig.~S2.

\subsection{Definition of the region of interest}

To define the different regions of interest on which we evaluated the data, we used the mean occupation and the mean spin imbalance per site.
In the loosely filtered case, we include all sites with $0.7< \langle \hat{n}_i \rangle<1.3$, effectively restricting the analysis region to the center of the cloud.
In the main text we also present results for a tighter density filter of $0.9< \langle \hat{n}_i \rangle<1.1$, additionally restricted to a single chain.
This chain was selected based on the highest nearest-neighbor spin correlations signal averaged over all data sets taken for interactions $U/t > 8$, for which we observed saturation of the spin correlations.\\

A second filter is required due to short-scale imperfections in the lattice potentials.
These are due to interference fringes in the projected lattice beams, which we minimized but did not removed entirely.
These fringes spoil the symmetry of the superlattice double wells locally, leading to a slightly biased splitting of the two spins in either direction on some sites.
To remove this effect from the data, we took into account only sites on which we detected a small enough spin imbalance (see Fig.~S3).
The precise superlattice phase control ensured a typical mean imbalance of $I=0.008(6)$. We filter sites that are incompatible with a splitting imbalance of less than $I=0.05$ on a $3.5\sigma$ level.
The position of the fringes drifts slowly on the timescale of days and we select the relevant sites per dataset, typically measured over several hours.
Distinct to the density based site selection, this last filtering is only required due to imperfections in the detection, not due to physical properties of the system.
The local occupation, imbalance and extracted spin correlators are shown in Fig.~S3 for the dataset measured at $U/t=12.6$.

\begin{figure}[t!]
\centering
\includegraphics[width=\columnwidth]{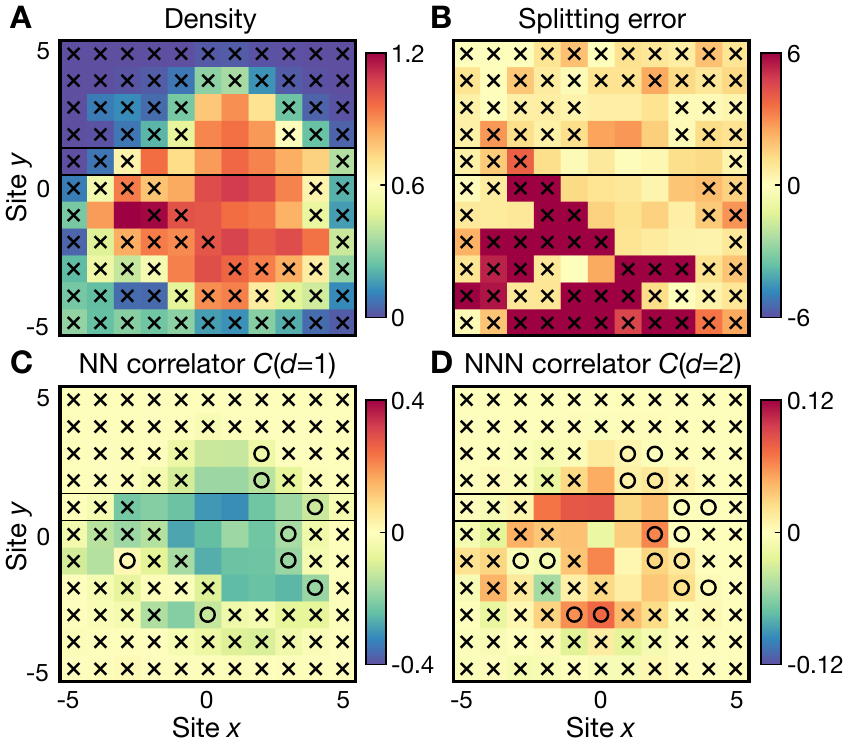}
\caption{
\textbf{Selection of the relevant lattice sites used in the statistical analysis.}
We define the region of interest for the statistical analysis based on the local density and the mean imbalance after spin splitting as described in the main text.
The figure shows four different local observables for the $U/t=12.6$ dataset.
Sites marked with an ``\textbf{x}'' are discarded due to the loose density or imbalance filter and the black box marks the tube considered in the tight filtered case.
\textbf{(A)} Density $\langle \hat{n}_i \rangle$ per lattice site.
\textbf{(B)} Splitting error in the spin resolved detection due to interference structures in the lattice beams.
Shown is the splitting error, that is, the mean detected spin imbalance per site after splitting, normalized to the root-mean-squared uncertainty of the mean of a binomial distribution.
We took only sites into account on which the splitting error was below $3.5$ standard deviations.
\textbf{(C)} Local measurement of the nearest neighbor spin correlator $C(d=1)$.
\textbf{(D)} Local measurement of the next-nearest neighbor spin correlator $C(d=2)$.
In (C) and (D) we draw the spin correlation only on the left partner of each pair. Sites accepted by the filter but without a partner site to the right are discarded and marked with an ``\textbf{o}''.
}
\end{figure}

\section{Calibration of Hubbard parameters}

\subsection{Lattice depth and tunneling calibration}

The generation of the lattice potentials via projection through the objective is described in our previous work~\cite{omran2015}.
We calibrated the depth of the various lattices via lattice modulation spectroscopy.
To this end, we used a spin polarized sample in the $\ket{1} = \ket{F=1/2,m_F=-1/2}$ state.
After the lattice was ramped to the desired value, we modulated
its intensity for $300\,$ms with an amplitude of $\pm3\%$.
We measured the number of transferred atoms from the ground to the second excited band versus the modulation frequency by counting the number of holes.
Finally, the tunneling rates were estimated from a band structure calculation for the measured lattice depths.

\begin{figure}[t]
 \centering
 \includegraphics[width=\columnwidth]{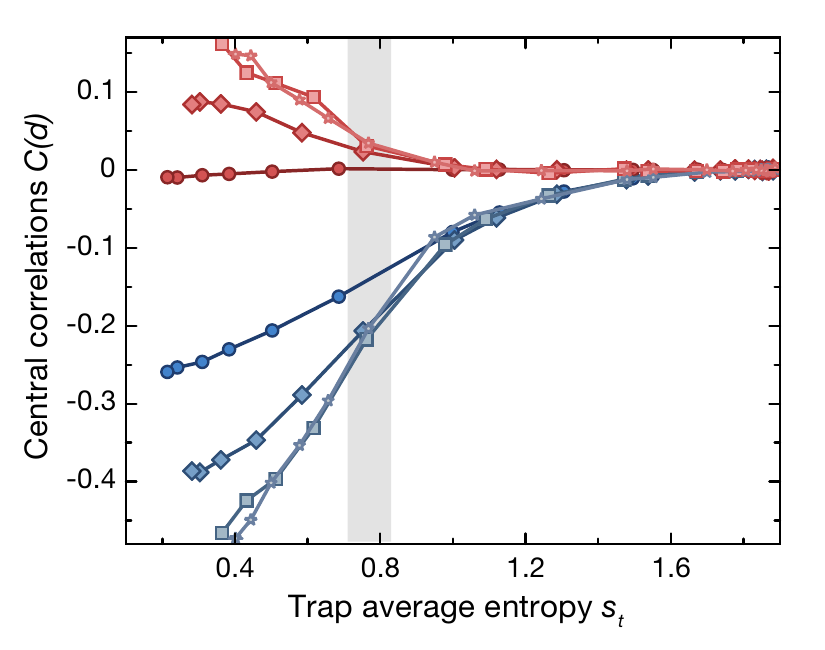}
 \caption{\textbf{Quantum Monte-Carlo results for the inhomogeneous system.}
  QMC calculations of the spin correlations $C(1)$ and $C(2)$ in the trap
  center as a function of the trap average entropy per particle $s_t$, with
  constant particles number $N \approx 22$ for $s_t \lesssim 0.9$. The blue
  lines correspond to $C(1)$ and the red lines to $C(2)$, while the symbols
  give the interaction strength: $U/t = 4$ (circle), $6$ (diamond), $8$
  (square), $10$ (star). The shaded area corresponds to the estimated trap
  average entropy per particle realized in the experiment. In contrast to the
  homogeneous case the correlations gradually increase when increasing the
  interaction strength $U/t$ from $4$ to $8$ for $s_t \lesssim 1$. Statistical
  errors are smaller than the symbols.
}
\end{figure}

\subsection{Calibration of the interaction strength}

\begin{table*}
  \centering
  \begin{tabular}{c|c|c|c}
    Magnetic field (G) & Scattering length ($a_B$) & Onsite interaction $U/h$ (Hz) & U/t \\
    \hline
    $529$ & $8  $ & $19 $ & $0.1 $ \\
    $573$ & $200$ & $463 $ & $3.7 $ \\
    $586$ & $272$ & $630 $ & $5.1 $ \\
    $598$ & $353$ & $820$ & $6.6 $ \\
    $611$ & $445$ & $1034$ & $8.4 $ \\
    $624$ & $550$ & $1279$ & $10.3$ \\
    $637$ & $671$ & $1560$ & $12.6$ \\
    $649$ & $810$ & $1883$ & $15.2$ \\
    $657$ & $904$ & $2101$ & $17.0$ \\
  \end{tabular}
  \caption{Summary of the interaction parameters used in the measurements.}
\end{table*}

The calibration of the interaction strength is based on the magnetic field tuned $s$-wave scattering length $a_s$, which has been precisely characterized~\cite{zuern2013a}.
The required magnetic field calibration was done by locating the narrow Feshbach resonance at $543.286\,$G~\cite{hazlett2012} via its associated atom loss feature in a spin balanced sample, which allowed for a calibration of the magnetic field to $0.2\,$G.
The onsite interaction U was then calculated using the Wannier function of the lowest band for the measured lattice depths.
Because of the large lattice spacing, we expect only small corrections to the interaction strength due to multi-band effects~\cite{buechler2010}.
Even at the largest scattering length $a_s=904\,a_B$ used in this manuscript, the ratio to the shortest lattice spacing is only $a_s/a_l = 0.04$.
In Table~S1 we summarize the relevant values of the magnetic field, the scattering length and the calculated interaction strength.

\section{Quantum Monte-Carlo calculations}

The numerical results presented in the manuscript build on the mapping between
the one-dimensional fermionic Hubbard model and a system of two hard-core bosonic species with
on-site interspecies interactions~\cite{Jordan1928}.
Path integral Monte Carlo simulations with worm-type updates~\cite{Prokofev1998}, here employed in the implementation of Ref.~\cite{Pollet2007},  
have a linear scaling in the system volume when  simulating the resulting bosonic model. The method overcomes critical slowing down for systems near a phase transition, and also allows to treat the trap efficiently.
The charge $C_c$ (atomic density) and spin density wave $C_s$ are both diagonal observables with respect
to the Fock basis $ \{ \vert \ldots, n_j ,\ldots \rangle \} $ used in the algorithm, and can readily be computed as
\begin{equation}
C_{c(s)}(i,j) = \frac{1}{4} \left< (n_\uparrow^i \pm n_\downarrow^i)  (n_\uparrow^j \pm n_\downarrow^j)  \right>.
\end{equation}
Here, the arrows distinguish between the two species of hard-core bosons, the upper (lower) sign refers to the charge $c$ (spin $s$) density wave, and $i$ and $j$ are site indices.

All the homogeneous simulations were carried out  in the grand-canonical
ensemble on chains of size $L=20$, which was checked to be already large enough
to rule out finite size corrections for the (local) quantities and the
parameters regimes we are interested in. In the experiment, the atom number as
well as the total magnetization in each tube fluctuates from shot to shot
justifying the use of the grand-canonical approach.
Note that in the canonical ensemble some quantities, for example, the nearest-neighbor
spin-spin correlator, show very strong finite size effects at half filling and
large values of $U/t$. For the system sizes and temperatures of interest here,
the values can differ up to $50\%$ from the grand-canonical values, the latter
being much closer to the thermodynamic limit.

The computation of the entropy per particle $s \equiv S/N k_B$ is usually more
cumbersome in Monte Carlo simulations. At infinite temperature it is
$s(\beta=0) = 2\, \ln(2)$ while at at zero temperature, $s(\beta=\infty) = 0$
(in this section, all energies are measured in units of the hopping $t$ and
$\beta=1/k_BT$). For intermediate temperatures,  the entropy is obtained by
numerical integration of the thermodynamic relation
\begin{equation}\label{eq:entropy_integral}
 S(\beta) = S(\beta_{\rm ref}) + \int_{\beta_{\rm ref}}^\beta
    \beta' \frac{dE}{d\beta'}\, d\beta'
\end{equation}
where $\beta_{\rm ref}$ is a reference inverse temperature for which
$s(\beta_{\rm ref})$ is known.
The total energy $E(\beta)$ is obtained from the Monte Carlo simulations and interpolated using a cubic
spline, which reduces the error.
The standard choice was $\beta_{\rm ref} = 0$, but we
crosschecked the results by using high temperature series 
expansion methods~\cite{PanWang1991} for $\beta \lesssim 0.01$ and
by repeating the integration procedure starting from $\beta_{\rm ref} =
\infty$, finding very good agreement between the different methods.
The error on $s$ is dominated by the uncertainty on $E(\beta)$, and was
quantified by bootstrapping the Monte Carlo energy samples and the integration
procedure.

Simulations of the trapped system were carried out with a harmonic confining
potential $V(r) = vr^2$ coupled to the particle density, effectively
modifying the local effective chemical potential, $\epsilon(r) = \mu - V(r)$,
cf.\  Eq.\ 1.
We set the trap so that $V=0$ at the central site of each chain and kept the
trapping amplitude fixed to $v = 0.05$.
The total entropy was computed in the same way as described for the homogeneous 
case but with $\beta_{\rm ref} = 0.01$ in Eq.\ \ref{eq:entropy_integral}.
This temperature is still sufficiently high that the use of the high temperature series in 
combination with the local density approximation remains justified to obtain the reference
entropy.
Similarly to the experimental conditions, the measurements of local observables 
were restricted to the sites at the bottom of the trap (sites $i=-5,...,5$ with 
respect to the center of the trap), where we verified that the average filling 
is close to one for large $U/t$.
Furthermore, we adjusted the chemical potential for each interaction strength
$U/t$ to keep the total number of particles in the trap approximately constant
($N \approx 22$ for $\beta \gtrsim 2$, corresponding to $s_t \lesssim 0.9$).
This results in the central region of the trap being slightly doped towards higher densities for low $U/t$.

\begin{figure}[t]
\centering
\includegraphics[width=\columnwidth]{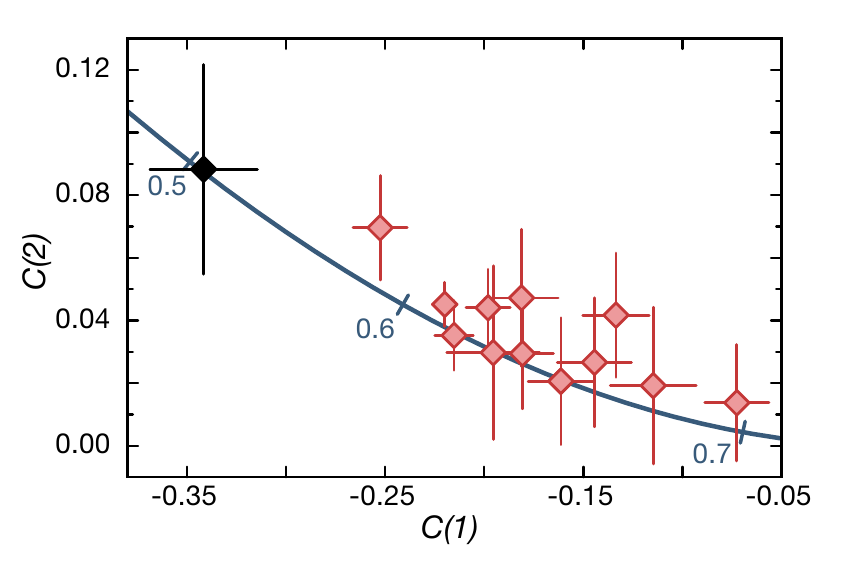}
\caption{
\textbf{Comparison between experimental spin correlations and quantum Monte-Carlo calculations in a homogeneous system.}
The red diamonds are experimental data for the spin correlations evaluated with the loose density filter. The data includes measurements for different interaction strengths in the saturated regime of the spin correlator $12.6<U/t<17$. Some of the measurements were taken for longer hold times before loading into the lattice, thus increasing the entropy. The black data point corresponds to the tighter filtered case described in the main text. The solid line is the quantum Monte-Carlo results for the spin correlators $C(1)$ and $C(2)$ for a homogeneous system at half filling and $U/t=12.6$ with different entropies $s$ indicated.
}
\end{figure}

\section{Entropy estimates}

As shown in Fig.~S4, the QMC simulations in the trap at constant total particle
number qualitatively explain the evolution of the spin correlations with
increasing interactions (cf. main text Fig.~3A). The entropy quoted in this
figure is the trap average entropy per particle $s_t$ and in the marked regime
around $s_t \approx 0.8$ the simulations reproduce the observed behavior of
increasing spin correlations between $U/t=4$ and the plateau region above
$U/t=8$. This indicates that the evolution of the trapped system is isentropic
also in the intermediate $U/t$ regime. Due to the imperfect knowledge of the
trap details in the experiment we did not attempt a more detailed comparison
here.

The spin correlator $C(d)$ provides a good relative thermometer since its
modulus monotonously decreases with temperature for a given lattice filling and
interaction strength. However, to obtain an absolute value for the temperature
or entropy, one needs to compare to a theoretical model. In Fig.~S5 we plot the
the correlators $C(1)$ and $C(2)$ for different data sets and compare them to
QMC calculations in a homogeneous system of varying entropy at half filling.
Based on the good agreement between theory and experiment one can infer the
entropy of the system by finding the Monte-Carlo dataset that best fits the
experimental correlators $C(d)$ for $d=1,2,3$.

\section*{Acknowledgements}
We acknowledge help by K. Kleinlein and M. Lohse during the setup of the experiment and financial support by MPG and EU (UQUAM, QUSIMGAS).




\bibliography{spincorr1d,blochgrouppapers}

\end{document}